\documentclass[preprint2]{aastex}

\usepackage[latin1]{inputenc}
\usepackage{array}

\shorttitle{Differentiation of 21 Lutetia}
\shortauthors{Formisano et al.}

\begin{document}

\title{The Onset of Differentiation and Internal Evolution: the case of 21 Lutetia}

\author{M. Formisano\altaffilmark{1,2}}
\email{michelangelo.formisano@iaps.inaf.it}
\author{D. Turrini\altaffilmark{1}, C. Federico\altaffilmark{1,3}, F. Capaccioni\altaffilmark{1}, M.C. De Sanctis\altaffilmark{1}}

\altaffiltext{1}{INAF-IAPS, Via del Fosso del Cavaliere 100, 00133 Roma (Italy)}
\altaffiltext{2}{University of Rome ``La Sapienza'', Piazzale Aldo Moro 5, 00185 Rome (Italy)}
\altaffiltext{3}{Dept. of Earth Science - University of Perugia, 06123 Perugia (Italy)}

\begin{abstract}

Asteroid 21 Lutetia, visited by the Rosetta spacecraft, plays a crucial role in the reconstruction of primordial phases of planetary objects.
Its high bulk density and its primitive chondritic crust \citep{Weiss11} suggest that Lutetia could be partially differentiated.

We developed a numerical code, also used for studying the geophysical history of Vesta \citep{Formisano12}, to explore several scenarios of internal evolution of Lutetia, differing in the strength of radiogenic sources and in the global post-sintering porosity.

The only significant heat source for partial differentiation is represented by $^{26}$Al, the other possible sources ($^{60}$Fe, accretion and differentiation) being negligible.

In scenarios in which Lutetia completed its accretion in less than 0.7 Ma from injection of $^{26}$Al in Solar Nebula and for post-sintering values of macroporosity not exceeding 30 vol.$\%$, the asteroid experienced only partial differentiation.

The formation of the proto-core, a structure enriched in metals and also containing pristine silicates, requires from 1 to 4 Ma: the size of the proto-core varies from 6 to 30 km.

\end{abstract}


\section{Introduction}

Asteroid 21 Lutetia plays an important role in the comprehension of the origin and evolution of planetary objects. As pointed out by \citet{Bottke}, the size of Lutetia allows its survival against impact disruption and that is why it preserves its original large scale structure. Data provided by the Rosetta spacecraft suggest an high bulk density (3400$\pm$300 kg m$^{-3}$) \citep{Patzold}: this evidence, combined to the primitive nature of the crust (carbonaceous or enstatite chondrites) \citep{Coradini11}, depicts a scenario in which 21 Lutetia experienced a partial differentiation with the formation of a metallic ``core" overlain by a primitive chondritic crust \citep{Weiss11}.

Geophysical and thermophysical history of Lutetia depends strongly on its initial composition and global macroporosity, which are greatly uncertain. Considerations on the global structure lead some authors to consider Lutetia as those asteroids with abundant fractures and joints \citep{Asphaug} for which the inferred macroporosity is in the range $\sim6-40\%$ \citep{Consolmagno08,Wilkinson}. Furthermore, all other asteroids having similar size (except 20 Massalia) are thought to have macroporosities of $>5-10\%$ and ranging up to $\sim80\%$ \citep{Consolmagno08}. A strict upper limit on Lutetia's macroporosity (i.e., $\sim52\%$) is provided by a model assuming that the entire asteroid is below a very thin chondritic surface layer and made up of pure iron.

\Citet{Weiss11} proposed three different scenarios for producing the high bulk density of Lutetia via partial differentiation. In the first scenario, primordial Lutetia has the same size as present and is initially undifferentiated. After radiogenic heating and subsequent internal melting, the metallic core and silicate mantle form with a decrease of the macroporosity and an increase of the bulk density. In the second scenario, primordial Lutetia has a larger radius and only a smaller volume fraction experiences the melting. The undifferentiated outer layer is removed by subsequent impacts while the bulk density increases. In the third scenario, the differentiation of Lutetia occurs early and at first the chondritic crust is not retained or it is removed by subsequent impacts: after this phase, chondritic debris are deposited on the asteroid.

In order to investigate which of these scenarios is the most plausible, we studied the thermal history of Lutetia. Before describing our model, however, we want to review the works already present in literature, as a wide range of thermal models of planetesimals assuming $^{26}$Al as the main heat source exists has been developed in recent years.

\Citet{Merk} analyzed the dependence of accretion rate on thermal evolution of planetesimal, concluding that accretion process had to be considered as long as the accretion time was not negligible respect to the half-life of the radionuclides providing energy. The authors also made use of Stefan law formulation in order to incorporate the effect of latent heat into thermal evolution, but they neglected the role of sintering and convection and they did not used a radiation boundary condition (i.e., they assumed a fixed surface temperature).

\Citet{Ghosh} also focused their attention on the importance of the accretion process in the thermal evolution of asteroids, by studying the case of 6 Hebe, showing the differences between instantaneous and incremental accretion cases. They used the model developed by \Citet{GMC} with a moving boundary condition and a radiation boundary condition. The authors concluded that incremental models provide a link between theoretical models of measurable quantities (i.e., peak temperatures, cooling rate, radioisotope closure times) in meteorites that were determined by their thermal histories.

\Citet{HS} incorporated convection in their thermal evolution models when the degree of partial melting exceeded 50 vol.$\%$. They analyzed the effects of sintering, whose onset was set at about 700 K, starting with a high porosity (50$\%$) and a low thermal conductivity. They considered instantaneous accretion and, as well as \citet{Merk}, they used a fixed boundary temperature. The main result of their work was to constrain the accretion of the parent bodies of differentiated meteorites to within the first 1.5 Ma, or at most 2 Ma, from CAIs. These planetesimals therefore formed before most chondritic parent bodies, which accreted at a later time.

\Citet{Mosko} studied the thermal consequences of melt migration and, in particular, they investigate how the redistribution of $^{26}$Al from the interior into a crustal layer would affect the thermal evolution of planetesimals. They considered the case of instantaneous accretion excluding sintering, convection and a radiation boundary condition, and they concluded that differentiation would be most likely for planetesimals larger than 20 km in diameter and that accreted within approximately 2.7 Ma from CAIs.

\Citet{SG} performed numerical simulations of the processes involving both aqueous alteration and planetary differentiation. They used asteroids of 100 and 270 km as representative cases to study, starting with high porosity and low thermal conductivity. They inserted the sintering and simulated the convection (at 50$\%$ of melting of silicates) by raising the thermal conductivity by three orders of magnitude. Their model assumes a fixed boundary temperature and does not take into account Darcy's law formulation for the segregation of silicates and melt. They concluded that a convective molten iron core is necessary to explain the remnant magnetism of the carbonaceous chondrite.

Finally, \Citet{ET} investigated the possibility that early radiogenic heating of planetesimals could create partially differentiated bodies with a primitive crust and magnetic core dynamos. The undifferentiated crust must be thick enough to prevent the majority of impacts from breaching it and persist through the internal magma ocean stage. They concluded that planetesimals accreting before about 1.5 Ma after CAIs are likely to fully differentiate through radiogenic heating, while bodies that accrete past about 1.5 Ma from CAIs would probably be characterized by a thick undifferentiated crust overlying a differentiated interior.

In the present work we apply the thermal code we developed to study the thermal history of Vesta \citep{Formisano12} in order to constrain the formation time, the size and the mass of the proto-core of Lutetia by varying the time-delay in the injection of  $^{26}$Al in the Solar Nebula (a similar approach is also presented in \citet{HS}, \citet{Mosko} and \citet{ET}) and the post-sintering macroporosity. We define as proto-core a structure enriched in metals, formed as a consequence of the metal percolation, but still containing pristine silicates.

As in \citet{HS} and \citet{ET}, we assume instantaneous accretion and solve the heat equation with radiogenic heating provided by $^{26}$Al and $^{60}$Fe.
As showed by \citet{Weiden}, the formation of a planetesimal having the same size as Lutetia requires about 10$^5$ year or less. Moreover, as pointed out by \citet{Mosko}, recent dynamical studies treating the turbulent concentration of small particles in proto-planetary disks \citep{Johansen07,Johansen09,Cuzzi} show that planetesimal can grow ``nearly instantaneously" in less than 100 yr to sizes of 100 km or larger. As a consequence, instantaneous accretion can be a reasonable approximation.

Similarly to \citet{Mosko}, we do not analyze the sintering phase due to the large uncertainties associated to the assumed initial porosities and to the corresponding thermal conductivities. 

We include the treatment of the latent heat based on the the Stefan law formulation, as in \citet{Merk}, in order to incorporate its effect on the evolution and we use Darcy's law model for metal percolation.

The developed scenarios will be useful to depict a reliable geophysical and thermal history and to constrain the internal structure of Lutetia. They also offer a complementary approach to the works present in literature, like the evidence for differentiation provided by gravitational features \citep{Vincent} or by hydrocode modeling of the largest impact crater on Lutetia \citep{Oklay}.

\section{The Model}
\subsection{Initial and Boundary Condition}

We consider 21 Lutetia as a spherical body of radius fixed to 50 km with initial homogeneous composition. We suppose that the object contains a metallic component ($Y$) (about 25 vol.$\%$) and a silicatic one ($X$) (about 75 vol.$\%$). This composition is similar to those of H and L class of ordinary chondrites. The post-sintering porosity ranges from 10 to 30 vol.$\%$. The initial temperature ($T_{0}$), which is also the temperature of the Solar Nebula in the surrounding of Lutetia, is fixed to 200 K \citep{Lewis}.
In Tab.\ref{tab:Parameters} we report the physical parameters used in our model.
We use a radiation boundary condition at the surface and a Neumann boundary condition at the center (heat flux equal to zero):
\begin{equation}
\begin{array}{l}
 T(r,t = 0) = T_0   \\
  \\
 \left[ {\frac{{\partial T}}{{\partial r}}} \right]_{r = 0}  = 0 \\
 \\
\left[ \frac{\partial T}{\partial r}\right]_{surf} = - \frac{\varepsilon \sigma}{K} \left(T_{surf}^4 - T_{0}^4\right),
 \end{array}
\end{equation}
where $T_{surf}$ is the temperature of the surface, $\varepsilon$ is the emissivity, $K$ is the thermal conductivity, and $\sigma$ is the Stefan-Boltzmann constant (see Tab.\ref{tab:Parameters}).

\subsection{Physical and Numerical Description}

We numerically study the thermal evolution of 21 Lutetia, solving the following coupled equations:
\begin{equation}\label{eq:heat}
{\left( {\rho c} \right)_m}\frac{{\partial T}}{{\partial t}} = \vec \nabla  \cdot \left( {{K_m}\vec \nabla T} \right) + H,
\end{equation}
and
\begin{equation}\label{eq:chim}
\frac{{\partial Y}}{{\partial t}} + \vec v \cdot \vec \nabla Y = 0.
\end{equation}
Eq.(\ref{eq:heat}) is the heat equation with the radiogenic heat source (H) \citep{Castillo07}, where $(\rho c)_m$ and $K_m$ are the overall heat capacity and the overall thermal conductivity, respectively. Eq.(\ref{eq:chim}) is the advection equation which controls the metal percolation in the silicatic matrix, with the migration velocity being a function of the permeability of the silicatic medium, of the density contrast between molten metals and solid silicates, of the gravitational acceleration and of the viscosity of the molten iron \citep{Yoshino04}.
As in \citet{GMC}, when melting temperature of Fe-FeS (1213 K) is reached, the percolation of the metals takes place through the silicatic matrix and the formation of the proto-core occurs. To take into account in a simple way the latent heat during phase transition, the specific heat is modified through the Stefan coefficient:
\begin{equation}\label{eq:Ste}
Ste =  \frac{L}{c}\frac{{d\chi }}{{dT}} = \frac{L}{c}\frac{1}{{T_{liq} - T_{sol}}},
\end{equation}
where $\chi$ is the degree of melting:
\begin{equation}\label{eq:chi}
\chi  = \frac{{T - T_{sol}}}{{T_{liq} - T_{sol}}},
\end{equation}
and so:
\begin{equation}\label{eq:c}
\bar{c} = c(1 + Ste),
\end{equation}
assuming a linear growth of $\chi$ with increasing temperature.

The numerical method is based on 1D, forward time-central space (FTCS), finite difference scheme in radial direction with Lax correction. A spatial grid ($\Delta r$) of 500 m is used.
To ensure the stability of our numerical approach, we use an \emph{adaptive} time increment according to the \emph{Courant-Friedrichs-Lewy} stability conditions for each of the physical processes (heat diffusion, metal percolation, radiation boundary condition) we consider in our work. Following \citet{Tok}, thermal conduction imposes the following critical time step:
\begin{equation}\label{eq:tstep}
\Delta t_{cond} = \frac{\left(\rho c\right)_m \Delta r^2}{2K_m}.
\end{equation}
In analogy with \citet{Tok}, we can define the following critical time step associated to the radiation boundary condition:
\begin{equation}\label{eq:trad}
\Delta t_{rad} = \frac{\left(\rho c\right)_m \Delta r T_{surf}}{\sigma\left( T^4_{surf} - T^4_0\right)}.
\end{equation}
Finally, while the percolation of metals is taking place, we need to solve also eq.(\ref{eq:chim}) and introduce a third critical time step:
\begin{equation}\label{eq:Courant}
\Delta t_{perc} = \frac{\Delta r}{v},
\end{equation}
where $v$ is the velocity of the metal percolation.
The Courant-Friedrichs-Lewy stability condition requires that the time step used in our model satisfies the following criterion:
\begin{equation}
\Delta t < min\left(\Delta t_{cond}, \Delta t_{rad}, \Delta t_{perc}\right).
\end{equation}
Therefore, at each temporal iteration of the program we select the actual time step basing on the minimum critical time step among those we computed. As the stability condition requires the actual time step to be lower than the critical one, we chose to use a value equal to 90$\%$ of the smallest critical time step as a compromise between the competing needs for stability and performances. So, our time step is defined as:
\begin{equation}
\Delta t = 0.9 \times min\left(\Delta t_{cond}, \Delta t_{rad}, \Delta t_{perc}\right).
\end{equation}
\section{Results}

We explored several scenarios characterized by different strength of the energy sources (the radiogenic heat due to the decay of $^{26}$Al) and values of post-sintering macroporosity (10, 20 and 30 vol.$\%$). The scenarios are labeled N0 (instantaneous accretion, $\Delta t_d = 0$), N1 ($\Delta t_d \simeq 0.3$ Ma) and N2 ($\Delta t_d \simeq 0.7$ Ma). The main results are showed in Tab.\ref{tab:Results}, in which we report the size, the time of formation and the mass of the proto-core (i.e., a structure enriched in metals and containing pristine silicates), and maximum temperature reached after 5 Ma. In all the cases we analyzed, the maximum degree of silicate melting ($\sim$10 vol.$\%$, corresponding to about 1450 K, \citet{Taylor1}) is reached only in a limited region of Lutetia. Since the degree of silicate melting is very low and, following \citet{HS}, the onset of convection should require that silicate melt fraction exceeds 50$\%$, we conclude that heat transport via convection is negligible. In fact, even in those regions where a significant temperature difference is present, the Rayleigh number does not overcome the critical values lying between 1000 and 2000.

In the case of post-sintering porosity of 10 vol.$\%$, we observe that in all scenarios Lutetia does not completely differentiate and only a proto-core forms (see the maximum temperature versus time profile of Fig.\ref{fig:Fig1}(f)). In Fig.\ref{fig:Fig1}(a), after 0.1 Ma, in the three scenarios the temperature is lower than the solidus temperature of silicates and the asteroid is homogeneously heated. After 0.5 Ma (see Fig.\ref{fig:Fig1}(b)), N0 enters in the melting temperature of silicates, while in N1 and N2 the temperature are lower than 1425 K.
After 1 Ma (see Fig.\ref{fig:Fig1}(c)), in N1 the temperature reaches the solidus temperature of silicates, while in N2 the values are still low. In Fig.\ref{fig:Fig1}(d), after 3 Ma, we observe a slight general increase of the temperature in all scenarios and then, after 5 Ma (see Fig.\ref{fig:Fig1}(e)), in N0 the temperature has almost the same value while in N1 and N2 we observe a general decrease. The proto-core size ranges from 25 to 31 km: the time of formation ranges from 1.2 to 2.2 Ma.

The general trend, for a porosity of 20 vol.$\%$, is similar to the previous case, but the values of temperature reached are lower because of the lower amount of material, and therefore energy sources, per unity volume (see Fig.\ref{fig:Fig2}(a)). As we can observe in Fig.\ref{fig:Fig2}(b), after 0.5 Ma, the temperatures overcome the liquidus temperature of metallic component in N0 and N1, while in N2 the temperature is in the window of melting of metals. After 1.5 Ma (see Fig.\ref{fig:Fig2}(c)), Lutetia is in the heating phase for all the scenarios and, after 3 Ma (see Fig.\ref{fig:Fig2}(d)), the general trend is the same. In Fig.\ref{fig:Fig2}(e), after 5 Ma, we observe a general decrease of the temperatures for N1 and N2. In Fig.\ref{fig:Fig2}(f) the maximum temperature versus time profile is reported. We observe that the maximum temperature is reached in the hottest scenario, characterized by instantaneous accretion (N0), and the time of formation of the proto-core ranges from 1.7 to 3.2 Ma while the proto-core size ranges from 14 to 21 km.

If we choose a value of porosity of 30 vol.$\%$, we can observe that, after a isothermal phase (see Fig.\ref{fig:Fig3}(a)) for all scenarios, at 0.5 Ma the temperature overcomes the liquidus melting temperature of metals in N0, while in N1 it is in the windows of melting of metals and, in N2, it is lower than 1213 K (see Fig.\ref{fig:Fig3}(b)). After 1.5 Ma (see Fig.\ref{fig:Fig3}(c)), N1 overcomes the liquidus temperature of metals and N2 enters in the windows of melting of metals. The general trend is the same after 3 and 5 Ma (see Fig.\ref{fig:Fig3}(d) and (e), respectively). In no scenario the temperature reaches the solidus temperature of silicates (see Fig.\ref{fig:Fig3}(f)). The formation of the proto-core occurs from 2.3 to 4 Ma. Maps of Fig.\ref{fig:Fig4} summarize the results obtained for the three cases of porosity explored.

In Fig.\ref{fig:Fig4}(a-b-c) maximum temperatures are reached in the middle region of the asteroid (from 30 to 40 km from the center) as a consequence of partial differentiation: in this region, in fact, there is less mass to be heated and so the temperatures are higher. In Fig.\ref{fig:Fig4}(d-e-f) we can see that the general trend is the same of Fig.\ref{fig:Fig4}(a-b-c) but the temperatures reach lower values than in the previous case because the porosity is increased (i.e., 20 vol.$\%$). In N0 the melting of silicates is possible, while in N1 and N2 only the melting of metals occurs. Fig.\ref{fig:Fig4}(g-h-i) show that high values of porosity (i.e., 30 vol.$\%$) prevent the reaching of the melting temperature of silicates and in particular, when the delay in the injection of $^{26}$Al is greater (i.e., 0.72 Ma), the melting of metals is possible only in a narrow region of Lutetia, ranging from 5 to 15 km.

Although we would expect an increase of temperatures with increasing porosity due to lowering of the thermal conductivity \citep{Opeil}, the dominant effect of the increase in porosity is the decrease of the concentration of the radioactive source ($^{26}$Al) linked to the silicates: therefore, the hottest scenarios are those characterized by the lowest porosity, as we can see in Tab.\ref{tab:Results}.

We also observe that in all the scenarios analyzed a surface layer ranging from 2 to 5 km is below the Curie temperature of carbonaceous chondrites (corresponding to about 873 K): this means that a remanent magnetization of this body is possible \citep{Richter}. The remanent magnetization of Lutetia has not a certain nature: it could be externally generated by external sources in the primordial Solar System \citep{Weiss10} or due to the presence of an internal core dynamo that could confirm the scenario of partial differentiation.

We can make a direct comparison between our temperature profiles and those of other models. Comparing our results with \citet{HS}, in case of 50 km body and instantaneous or late accretion (0.75 Ma), maximum temperature reached (about 8000 and 4000 K, respectively) are very different from our values (in the hottest scenario of current paper, i.e., N0 and 10 vol.$\%$ of porosity, we obtain 1455 K). Our thermal profiles also show a maximum in the ``mantle", due to the differentiation and the affinity of $^{26}$Al with silicates. These global differences probably depend on the different methodology and initial conditions adopted in the two works. The authors concluded that the thin crust that formed is destined to be processed by the magma flowing beneath due to small impacts or convective drag. On the contrary, \citet{ET} concluded that undifferentiated chondritic crust survive through the internal magma ocean phase. If we define the formation of magma ocean at 50 vol.$\%$ of silicate melting, the small degrees of silicate melting reached in our scenarios prevent this situation and preserve the primitive unmelted crust. \citet{SG} also showed the existence of a chondritic crust even but they assumed a fixed boundary temperature instead of a more realistic radiation boundary condition, so that the extent of the their chondritic crust was likely overestimated. Since the methodology assumed in \citet{Mosko} is similar to that of our work, maximum values of temperature are more compatible with our owns than those of \citet{HS}, even if global profiles are different.

\section{Summary and Conclusions}

Observational data do not provide stringent constraints about the internal structure. Currently we know that Lutetia possesses a chondritic crust (carbonaceous or enstatitic) and its high bulk density has been interpreted as in indication of the presence of a metallic core. The results of our model suggest that partial differentiation is possible: in fact, the maximum degree of silicate melting is about 10 vol.$\%$ in a limited region of the ``mantle". This is consistent with the scenarios proposed by \citet{Weiss11}, if the current macroporosity (10 - 30 vol.$\%$) is the same as the post-sintering one. In all scenarios only the formation of a proto-core, a structure enriched in metals, occurs. The proto-core formation takes from 1 to about 4 Ma and proto-core size ranges from 6 to 30 km. The relative proto-core mass ranges from about 1 to about 36 $\%$ of the total mass. Our results suggest that the accretion time {of Lutetia should not exceed 0.7 Ma from CAIs and the post-sintering macroporosity does not exceed 30 vol.$\%$.

In all scenarios we considered, a primitive, undifferentiated crust survives thermal evolution \citep{ET,SG} and it possibly could be reduced by subsequent impacts.

A possible remanent magnetization is retained, if we assume a carbonaceous surface composition, since in all the scenarios \textbf{we} explored a surface layer of about 2-5 km is below the Curie temperature.

The main source of energy is represented by $^{26}$Al while the contribution due to $^{60}$Fe and other possible sources (e.g. accretion and differentiation processes) is negligible.

\acknowledgments

We wish to thank Guy J. Consolmagno and an anonymous referee for their helpful comments and Romolo Politi for his numerical analysis assistance. M.F. thanks his friend Demetra De Cicco for the revision of the text. The computational resources used in this research have been supplied by INAF-IAPS through the project HPP-High Performance Planetology.

\begin{table*}[!H]
\resizebox{1\textwidth }{!}{
\begin{tabular}[c]{lccc}
  \hline
  \textbf{Quantity}  & \textbf{Value} & \textbf{Unit} & \textbf{Reference}\\
  \hline
  & & & \\
  Final primordial radius ($R$) & 50$\times10^{3}$ & $m$ & \citet{Weiss11}\\
  Density of metal ($\rho_{met}$) & 7800 & $Kg$ $ m^{-3}$ & \citet{Sramek}\\
  Density of silicate ($\rho_{sil}$) & 3200 & $Kg$ $ m^{-3}$ & \citet{Sramek}\\
  Specific heat of metal (solid) ($c_{met,sol}$) & 600 & $J Kg^{-1} K^{-1}$ & \citet{Sahijpal}\\
  Specific heat of metal (liquid) ($c_{met,liq}$) & 2000 & $J Kg^{-1} K^{-1}$ & \citet{Sahijpal}\\
  Specific heat of silicate(solid) ($c_{sil,sol}$) & 720 & $J Kg^{-1} K^{-1} $ & \citet{Sahijpal}\\
  Specific heat of silicate (liquid) ($c_{sil,liq}$) & 720 & $J Kg^{-1} K^{-1}$ & \citet{Sahijpal}\\
  Latent heat of metal ($L_{met}$) & 270 & $ KJ $  $ Kg^{-1}$ & \citet{GMC}\\
  Latent heat of silicate ($L_{sil}$) & 400 & $ KJ $  $ Kg^{-1}$& \citet{GMC}\\
  Metal solidus ($T_{sol}^{met})$ & 1213 & K & \citet{GMC}\\
  Metal liquidus ($T_{liq}^{met})$ & 1233 & K & \citet{GMC}\\
  Silicate solidus ($T_{sol}^{sil}$) & 1425 & K & \citet{Taylor1}\\
  Silicate liquidus ($T_{liq}^{sil}$) & 1850 & K & \citet{Taylor1}\\
  Thermal conductivity of metal ($K_{met}$) & 50 & $W$ $m^{-1}$ $K^{-1}$ & \citet{Sramek}\\
  Thermal conductivity of silicate ($K_{sil}$) & 3 & $W$ $m^{-1}$ $K^{-1}$ & \citet{Sramek}\\
  Initial metal volume fraction ($Y$)& $25\%$&  & \\
  Initial silicate volume fraction ($X$) & $75\%$& & \\
  Post-sintering porosity ($\phi$) & $10\%$ - $30\%$ &  & \\
  Temperature of Solar Nebula ($T_0$) & 200 & $K$ & \citet{Lewis} \\
  Stefan-Boltzmann constant ($\sigma$) & 5.67$\times 10^{-8}$ & $W$ $m^{-2}$ $K^{-4}$ &\\
  Emissivity ($\varepsilon$) & 1 &  &\\
  Half-life of $^{26}Al$ ($\tau_{Al}$) & $0.717$ & $Ma$ & \citet{Castillo09}\\
  Specific heat production of $^{26}Al$ & $0.355$ & $W Kg^{-1}$ & \citet{Castillo09}\\
  Initial isotopic abundance of $^{26}Al$ in ordinary chondrites ($[^{26}Al]_0$)& $6.20\times10^{-7}$ & $ppb$ & \citet{Castillo09}\\
  Half-life of $^{60}Fe$ ($\tau_{Fe}$) & $2.62$ & $Ma$ & \citet{Rugel}\\
  Specific heat production of $^{60}Fe$ & $0.068\div0.074$ & $W Kg^{-1}$ & \citet{Castillo07}\\
  Initial isotopic abundance of $^{60}Fe$ in ordinary chondrites ($[^{60}Fe]_0$)& $(22.5\div225) \times10^{-9}$ & $ppb$ & \citet{Castillo07}\\
  & & & \\
  \hline
\end{tabular}}
\centering
\caption{Physical parameter values used in this work.}\label{tab:Parameters}
\end{table*}

\begin{table*}[H]
\centering
\begin{tabular}[c]{|c|c|c|c|c|}
  \hline
  &\textbf{Size} [km] & \textbf{$\Delta t_{core}$} [Ma] & \textbf{$M_{core}$ $[\% M_{tot}$]} & \textbf{$T_{max}$} \\
  \hline
  \textbf{N0} &&&&\\
  $\phi = 10$ vol.$\%$ & 31 & 1.2 & $\simeq36$ & 1455\\
  $\phi = 20$ vol.$\%$ & 21 & 1.7 & $\simeq11$ & 1443\\
  $\phi = 30$ vol.$\%$ & 13 & 2.3 & $\simeq3$ & 1338\\
  \hline
  \textbf{N1} &&&&\\
  $\phi = 10$ vol.$\%$ & 28 & 1.6 & $\simeq27$ & 1448\\
  $\phi = 20$ vol.$\%$ & 18 & 2.3 & $\simeq7$ & 1357\\
  $\phi = 30$ vol.$\%$ & 11 & 2.8 & $\simeq2$ & 1281\\
  \hline
  \textbf{N2} &&&&\\
  $\phi = 10$ vol.$\%$ & 25 & 2.2 & $\simeq19$ & 1387\\
  $\phi = 20$ vol.$\%$ & 14 & 3.2 & $\simeq3$ & 1284\\
  $\phi = 30$ vol.$\%$ &  6 & 3.6 & $<1$ & 1239\\
  \hline
\end{tabular}
\caption{Summary of scenarios. Size, time formation and mass of the proto-core and maximum temperature reached after 5 Ma are reported.}\label{tab:Results}
\end{table*}

\begin{figure*}[H!]
\includegraphics[width=0.9\textwidth]{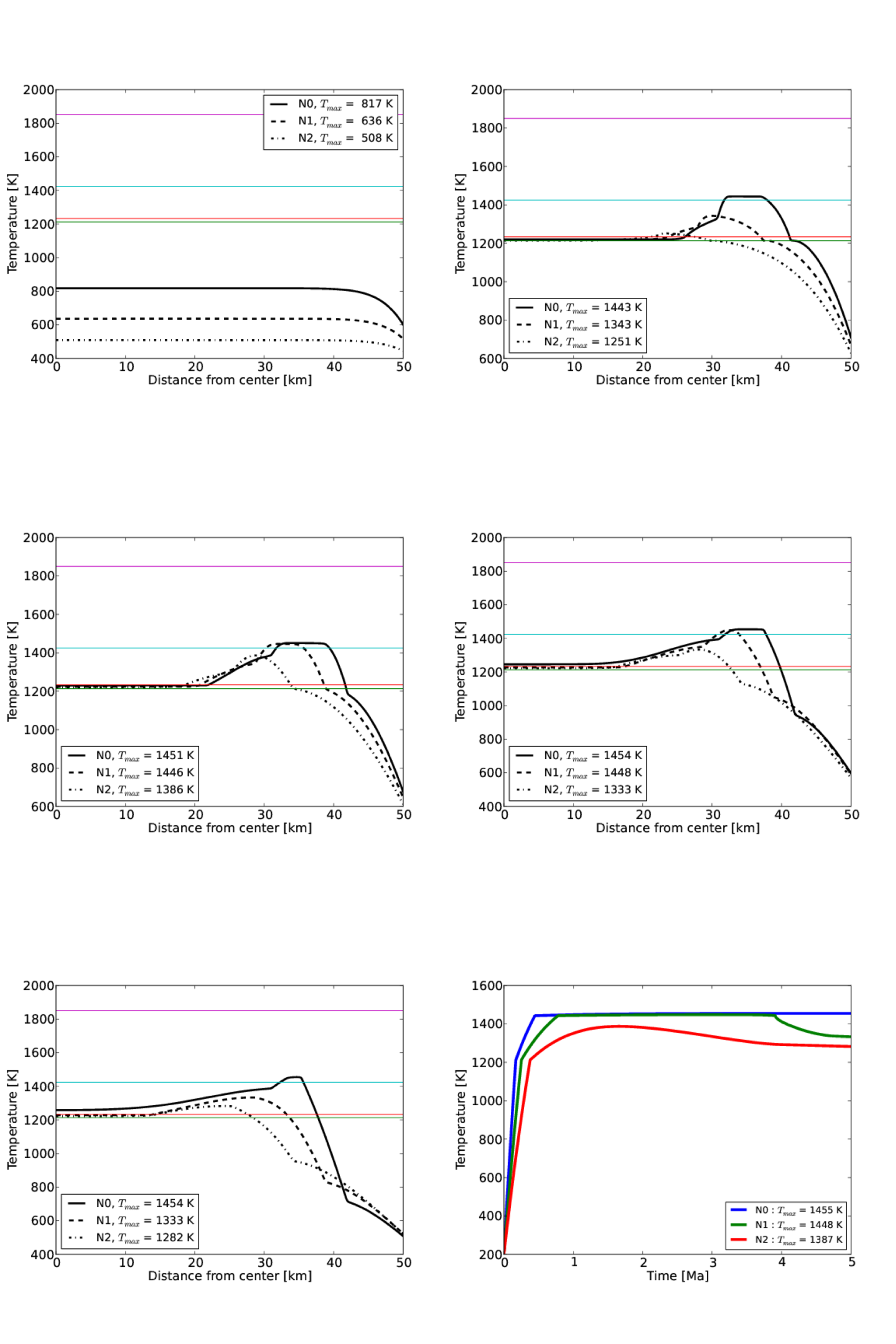}
\caption{From top to bottom, from left to right, plots are: temperature vs distance from center at 0.1
Ma (a), at 0.5 Ma (b), at 1.5 Ma (c), at 3.0 Ma (d), at 5.0 Ma (e) and maximum temperature profile vs
time (f), for $\phi$= 10 vol.$\%$.}\label{fig:Fig1}
\end{figure*}

\begin{figure*}[H!]
\includegraphics[width=0.9\textwidth]{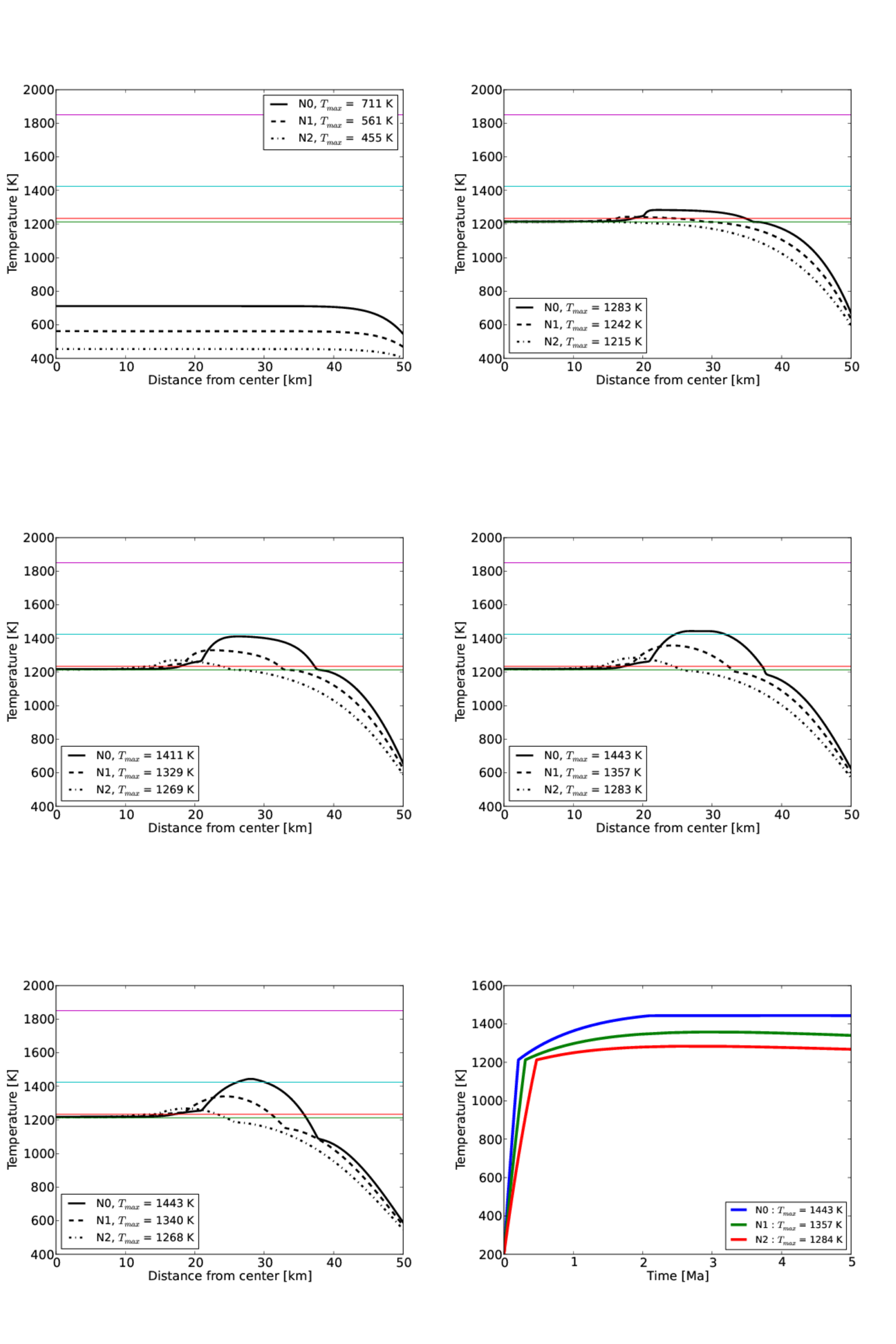}
\caption{From top to bottom, from left to right, plots are: temperature vs distance from center at 0.1
Ma (a), at 0.5 Ma (b), at 1.5 Ma (c), at 3.0 Ma (d), at 5.0 Ma (e) and maximum temperature profile vs
time (f), for $\phi$= 20 vol.$\%$.}\label{fig:Fig2}
\end{figure*}

\begin{figure*}[H!]
\includegraphics[width=0.9\textwidth]{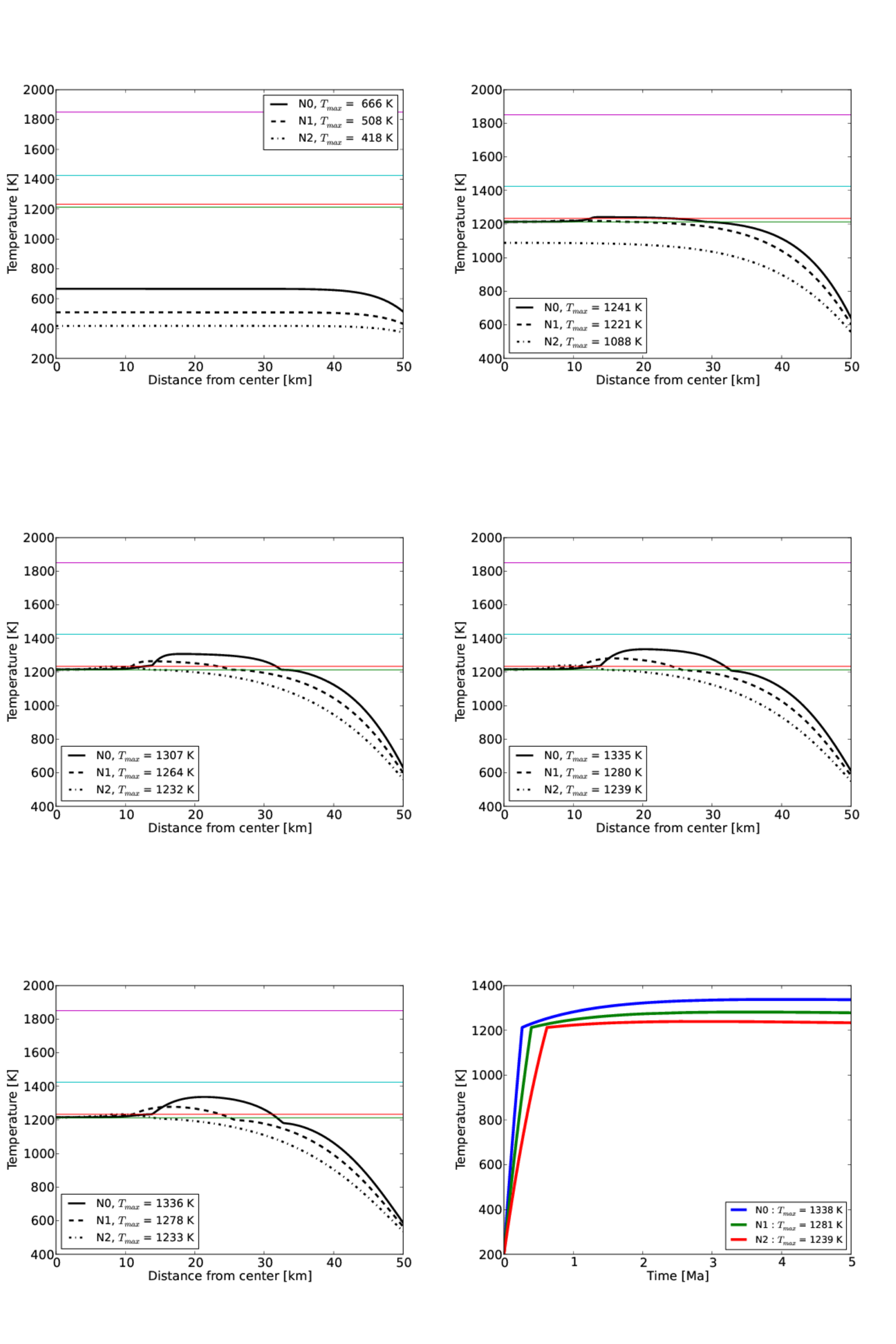}
\caption{From top to bottom, from left to right, plots are: temperature vs distance from center at 0.1
Ma (a), at 0.5 Ma (b), at 1.5 Ma (c), at 3.0 Ma (d), at 5.0 Ma (e) and maximum temperature profile vs
time (f), for $\phi$= 30 vol.$\%$.}\label{fig:Fig3}
\end{figure*}

\begin{figure*}[H!]
\includegraphics[width=\textwidth]{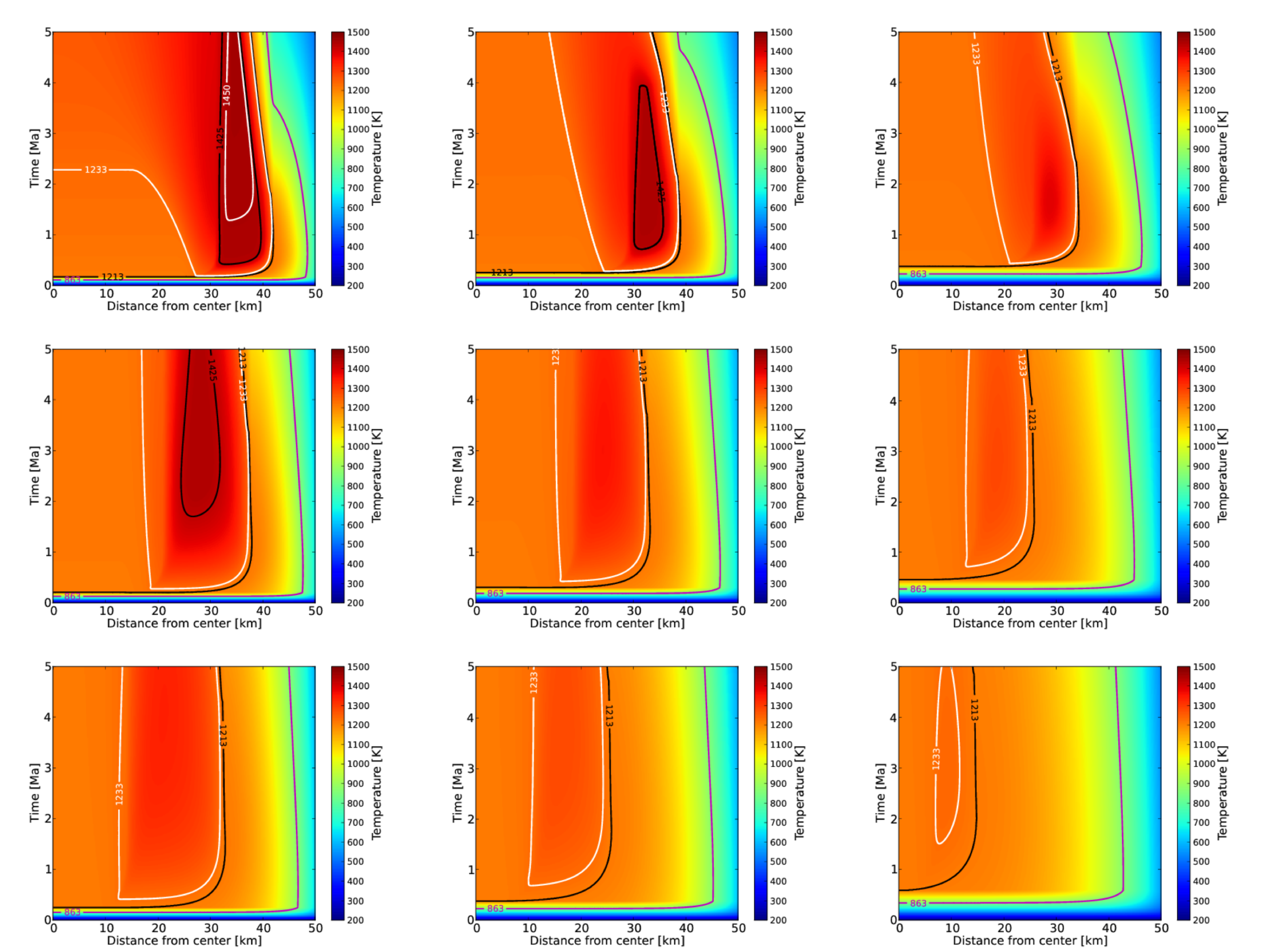}
\caption{In the top line plots are: thermal history maps of N0 (a), N1 (b) and N2 (c) for $\phi$ = 10 vol.$\%$; in the middle line plots are: thermal history maps of N0 (d), N1 (e) and N2 (f) for $\phi$ = 20 vol.$\%$; in the bottom line plots are: thermal history maps of N0 (g), N1 (h) and N2 (i) for $\phi$ = 30 vol.$\%$. Magenta isoline represents the Curie temperature of carbonaceous chondrites.}\label{fig:Fig4}
\end{figure*}

\end{document}